\documentclass[12pt,a4paper]{article}

\usepackage[german, english]{babel}
\usepackage{a4wide}
\usepackage{amsmath}
\usepackage{amssymb}
\usepackage{ifthen}
\usepackage{epsfig,color}

\usepackage{bm}
\usepackage[T1]{fontenc}
\usepackage{graphicx}

\begin{document}
	\begin{titlepage}

\title{Hidden Vacuum Rabi oscillations: Dynamical Quantum Superpositions of On/Off Interaction between a Single Quantum Dot and a Microcavity}
\author{A.\ Ridolfo $^{1}$, R.\ Stassi$^{1}$, O. Di Stefano $^{1}$
\\
{\it {\scriptsize
			$^1$
Dipartimento di Fisica e Scienze della Terra, Universit\`{a} di Messina,
 Viale Stagno D'Alcontres 31 I-98166 Messina, Italy}}\\
{\scriptsize{ E-mail: odistefano@unime.it}}
	}

\maketitle

\begin{abstract}
{
We show that it is possible to realize quantum superpositions
of switched-on and -off strong light-matter interaction in a single quantum dot- semiconductor microcavity system.
Such superpositions enable the observation of counterintuitive quantum conditional dynamics effects.
Situations are possible where cavity photons as well as
the emitter luminescence display exponential decay but their joint detection probability exhibits  vacuum Rabi oscillations.
Remarkably, these quantum correlations are also present in the  nonequilibrium steady state spectra of such coherently driven
dissipative quantum systems.
}
\end{abstract}

\bigskip

\noindent{Keywords:~~Rabi oscillations, Quantum Superposition,Nanostructures, Light-matter interaction
\newline
}

\end{titlepage}


%

\section{Introduction}
Strong coherent
interactions between quantum emitters  (QEs) and photons
play a key role in several schemes for
the realization of quantum logic gates  \cite{Pellizzari,Zubairy,Monroe} and quantum networks \cite{Cirac, Ritter}.
A quantum phase gate operating on quantum bits carried by a single Rydberg atom
and a zero- or one-photon field in a high-Q cavity has been demonstrated \cite{Haroche}.
It has been shown that a cavity with a single trapped
atom, can be exploited to realize scalable, fully functional photonic quantum computation \cite{PRLKimble}.
The implementation of quantum information processing with cavity quantum electro-dynamical (CQED) systems
requires full coherent control of superpositions of cavity-coupled and -uncoupled atomic levels \cite{Monroe,Haroche, PRLKimble}.
The superposition principle is at the heart of the most
intriguing features of  quantum mechanics.

A QE-microcavity system enters the strong coupling regime provided the coherent coupling overwhelms
the dissipative processes.
In this case  a striking change in the system dynamics from the usual irreversible spontaneous emission to a reversible exchange of energy between
the emitter and the cavity mode can be observed \cite{Haroche}. As shown in Ref.\cite{Stassi2012} for a three-level atom, if the emitter is set into a superposition state of one state strongly coupled with
the cavity mode and one other state which is uncoupled,  it is possible to prepare a quantum system which is ``suspended'' between two completely different time evolutions. When a measurement is performed, only one of these possible dynamics is actualized.

Here we extend previous work on three-level atoms into a superposition of two different dynamical states \cite{Stassi2012}. Specifically,
i) we show that it is possible to prepare quantum superpositions of on/off strong coupling between a microcavity and a single quantum dot (QD);
ii) we demonstrate that these superposition effects can also be observed in nonequilibrium steady state spectra of such
dissipative quantum systems under continuous-wave excitations;
iii) we calculate the joint probability of detecting a cavity photon at time $t_1$ and a free photon from the decay of the exciton level at  $t_2$ which shows intriguing time-dependent correlation effects and detection-conditioned dynamics.  Surprisingly, this two-times correlation function shows that the presence of a subsequent detection event affects the previous photonic dynamics, although causality is not (of course) violated.


Moreover we show that the resulting system dynamics, ``suspended'' between  two different time evolutions,
can be directly probed by comparing ordinary and coincidence  photodetection.
Such dynamics superpositions enable novel counterintuitive nonclassical effects. Situations are possible where cavity photons as well as
the emitter luminescence display a weak-coupling exponential decay but their joint detection probability displays  a strong-coupling dynamics.
Remarkably we find that, these quantum correlations are also present in the  nonequilibrium steady state which is reached under
the influence of  coherent continuous-wave driving and dissipation.
In the last years  the strong coupling of semiconductor self-assembled
QDs to single modes of monolithic optical cavities was achieved \cite{Reith,Yoshie} opening
the possibility of  realizing quantum information tasks in solid-state CQED systems.
The quantum nature of  such strongly coupled exciton-photon system has been proved \cite{Imamoglu}.
Moreover, substantial progress has been made demonstrating
the capability of a single QD coupled to a photonic crystal
microcavity to control phase modulation of light at single photon level \cite{Faraon}.
It has  been shown that a non-adiabatic full time control of the strong light-matter interactions of cavity embedded single QE can be realized in principle \cite{Ridolfo}.
Strong coupling in a single QD-semiconductor microcavity system was achieved by coupling the optical mode of a photonic crystal or a micropillar microcavity to the excitonic transition of the QD.
In order to  realize on/off quantum superpositions we propose, instead, to tune the mode energy of the microcavity to the biexciton-exciton (XX-X) transition of the QD.
A similar four-level biexciton-exciton cascade configuration was theoretically investigated in order to analyze the simultaneous formation and stable propagation of slow optical soliton pairs in semiconductor QD \cite{Yang2011}.
Two electron-hole pairs (a biexciton) trapped in a QD recombine radiatively through a frequency-nondegenerate two-photon  cascade \cite{Young}.
In a first step, the biexciton decays into one of the two possible optically active single-exciton states (see Figure \ 1a)
by emitting an XX photon, followed by an excitonic decay to the dot ground state (G) by emitting a second X photon with typically somewhat different energy \cite{Moreau}.
Recently an ultrabright source of entangled photon pairs has been obtained by coupling an optical cavity in the form of a {\em photonic
molecule} to a single QD \cite{Bloch}.  The lowest energy photonic mode is resonantly coupled to the XX-X(Y) transition and the other to the X(Y)-G  which has a transition energy of about 4 meV higher than the XX-X(Y) transition. In this way the Purcell effect is fully effective for both the transitions greatly improving the extraction efficiency of entangled pairs.

\section{The model}

We consider a single mode microcavity polarized along the $\hat {\bf x}$ (horizontal) direction \cite{PRLIma} resonantly coupled with the XX-X transition. The cavity mode polarization greatly suppresses the decay channel XX$\to$ Y, being Y-excitons coupled to the $\hat {\bf y}$ polarized photons.
The cavity will be  also coupled with the transition X-G, although the cavity mode is detuned with respect to this transition by  $\Delta \approx 4$ meV (much larger than the cavity linewidth).
The system is driven by two $\hat {\bf x}$-polarized pulsed or continuous wave laser beams (see Figure \ 1).
One beam excites resonantly the cavity mode, the other is resonant with the X-G transition.
Other schemes based on circularly polarized beams could also be employed. However the present configuration works even when the
fine-structure splitting cannot be neglected.
\begin{figure}
	\centering
	\includegraphics[scale=0.6]{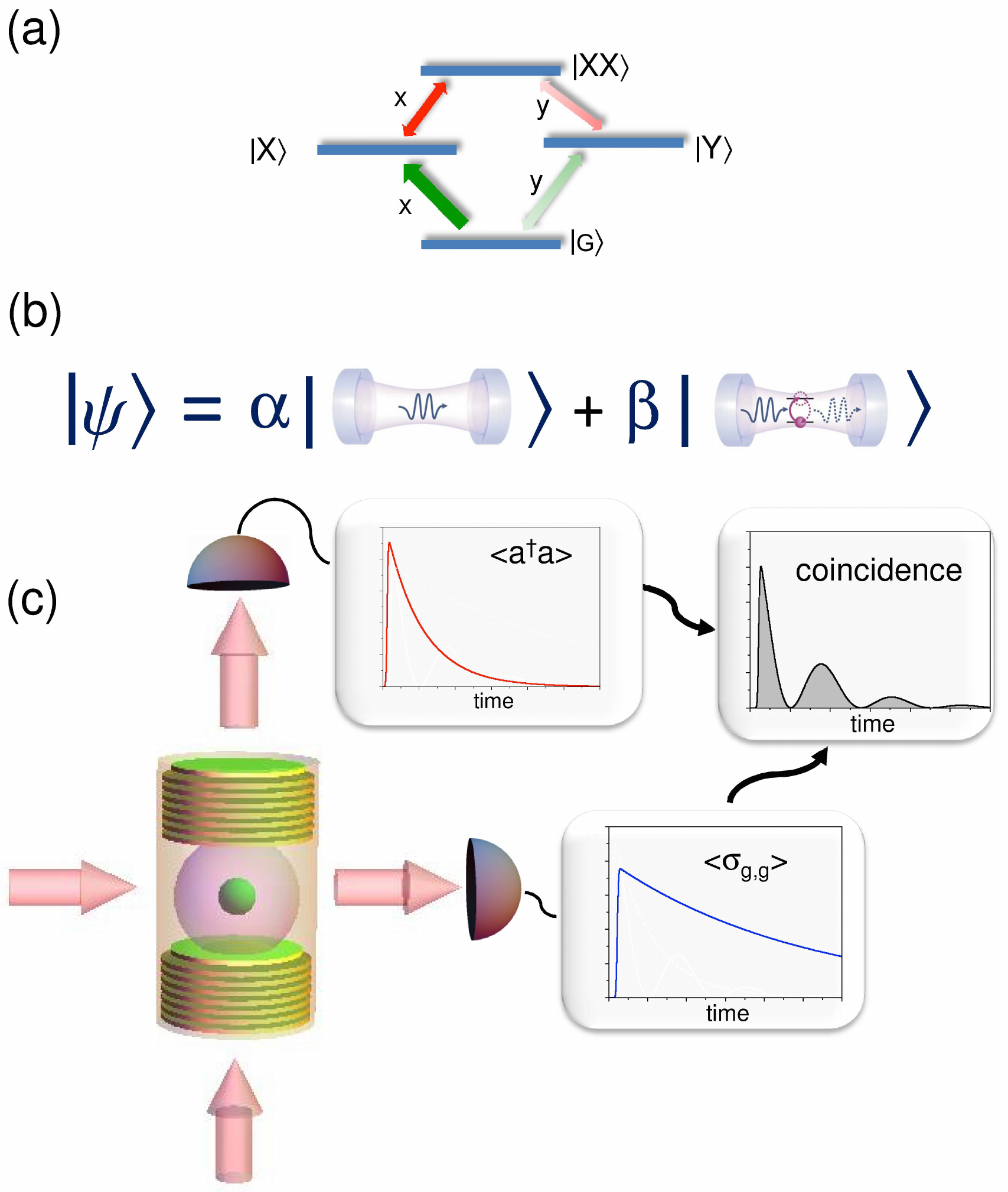}
	\caption{(Color online). (a) Sketch of the radiative cascade in a single QD. See main text for
		nomenclature. (b) Schematic description of the quantum superposition between
		a cavity strongly coupled with a two-level atomic-like transition with
		photons coherently absorbed and re-emitted at rate g and a bare cavity (without active material excitation).
		(c) Sketch of the microcavity embedded QD and of the excitation and detection schemes: quantum superposition (b) enables situations where cavity photons as well as the emitter luminescence display exponential decay but their joint detection probability exhibits vacuum Rabi oscillations.} 
\end{figure}
Before describing the detailed model and supporting calculations, first we summarize the basic ideas of our scheme. We consider only h-polarized pulses so that the state $| {\rm Y} \rangle$ is not excited.
At initial time the cavity is empty (zero photons) and the QE is described by
the state $| {\rm G} \rangle$.
We point out that the full width at half maximum (FWHM) of the pulse feeding the cavity is chosen less than the Rabi period ($FWHM<< \pi/g$). As consequence,  the cavity  is populated with photons before the energy transfer to the QE starts. The latter process, at the end, allows the coherent excitation cavity.
Then an h-polarized control pulse  of frequency $\omega_{\rm x}$ ($\omega_{\rm g}$ is set to zero) resonant with the $| {\rm G} \rangle \to | {\rm X} \rangle$ transition  determines a rotation of the Bloch sphere $\{| {\rm G} \rangle,\, | {\rm X} \rangle \}$.  After creation of the GX superposition, the cavity is excited by a weak resonant coherent pulse. Just after  the pulse arrival, the state becomes $| \psi'(t) \rangle = |\alpha \rangle ( a_{\rm G}| {\rm G} \rangle + a_{\rm X} | {\rm X} \rangle)$, being $|\alpha \rangle \approx | 0 \rangle + \alpha| 1\rangle$ a coherent photon state.
In fact, when the cavity is first excited by a weak probe beam	the levels coupled with the cavity  are unpopulated, hence
	cavity photons are not able to couple to the X-XX transitions.
	As a result, cavity photons start decaying exponentially
	according to the cavity decay time. As soon
	as the first control pulse excites the cavity
	the system switches to the strong-coupling regime, as
	witnessed by the presence of vacuum Rabi oscillations \cite{DiStefano2011}.
	 From now on  the two states composing the linear superpositions obey completely different dynamics: $|\alpha \rangle | {\rm G} \rangle$ is a stationary state (except for the cavity losses), while the ket $ |\alpha \rangle | {\rm X} \rangle$ evolves under the influence of the strong light-matter interaction which induces photon absorption and re-emission processes before dissipation becomes effective. The evolution of the
total quantum state can be schematically visualized as $| \psi(t)\rangle = a_{\rm X} | {\rm On}(t) \rangle + a_{\rm G} | {\rm Off}(t) \rangle$ (see Fig.\ 1b).
We have defined  $| {\rm Off}(t) \rangle=|\alpha \rangle | {\rm G} \rangle $ that, as observed above, is a stationary state. In addition, we have $| {\rm On}(t) \rangle_{t=0}=|\alpha \rangle | {\rm X} \rangle$. Such state evolves  as: 
$
| {\rm On}(t) \rangle= \sqrt{1- \alpha^2}|0\rangle |X\rangle + \alpha \left[\cos(gt)|1\rangle |X\rangle+\sin(gt)|0\rangle |XX\rangle\right]
$ \cite{Haroche}.
If $a_{\rm X}=0$ we have $| \langle a^\dag a \rangle= |\alpha|^2$ and as expected the system is in a stationary state; if $a_{\rm G}=0$ 
 $| \langle a^\dag a \rangle= |\alpha|^2 \cos^2(gt)$ giving rise to Rabi oscillations. For the general case ($a_{\rm X} \neq0$ and $a_{\rm G} \neq0$) we obtain a mixed behaviour and the  the system is suspended between two different dynamics.
Once 
the coupled quantum system is in this superposition state, probing it (e.g. feeding the cavity with a weak probe pulse), will give rise to an output signal with hybrid strong-weak
coupling properties which can be probed experimentally. Moreover
photons spontaneously emitted into free space from the decay of the exciton level, can be detected and collected in addition to  output cavity photons. At last, through a coincidence measurement, it is possible to post-select  only cavity photons in the strong coupling regime.

\begin{figure}
	\centering
	\includegraphics[scale=0.5]{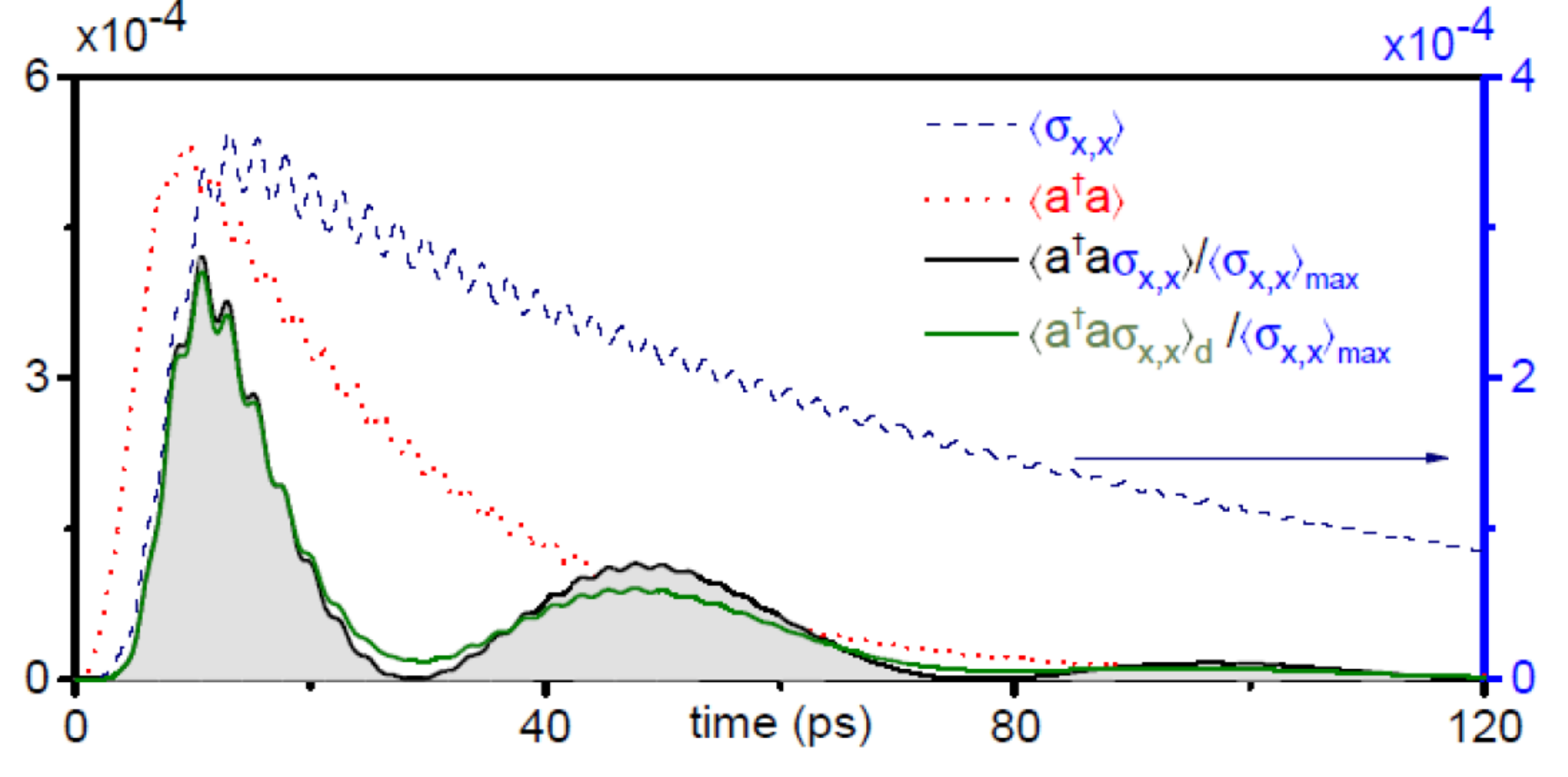}
	\caption{(Color online). Time evolution of cavity photons $\langle a^\dag a \rangle$, of spontaneously emitted photons from the exciton level
		$\propto \langle \sigma_{\rm x, x}\rangle$, and the coincidence rate of detecting one cavity photon and one photon from the spontaneous transition $| {\rm X} \rangle \to |{\rm G} \rangle$ which is proportional to $I_{\rm xc} = \langle a^\dag a\, \sigma_{\rm x, x}\rangle$ (filled curve) after the arrival of two weak optical pulses exciting resonantly both the cavity and the $| {\rm G} \rangle \to |{\rm X} \rangle$ transition; while the first two expectation values display almost exponential irreversible decay, the third exhibits vacuum Rabi oscillations. The Fig. also displays  $I_{\rm xc}$ calculated including pure dephasing.
		For the adopted parameters see text. }
\end{figure}
In the absence of losses the optically pumped quantum system depicted in Figure \ 1 is described by the Hamiltonian
\begin{eqnarray}
H &=&
\sum_{j= {\rm x}, {\rm xx}}
	\omega_j\, \sigma_{j,j}+
	\omega_{\rm c}\, {a}^{\dag} a
	+ \left[  g(a^\dag \sigma_{\rm x, xx}  +  a^\dag \sigma_{\rm g, x}) \right. \nonumber \\
	  &+& \left. {\cal E}_{\rm c}(t) ( \sigma_{\rm x, g} +\sigma_{\rm xx, x} )+ {\cal E}_{\rm in}(t)a^\dag + {\rm H.c.} \right]\, ,
\label{Hamiltonian}
\end{eqnarray}
where the Bosonic operators $a$ ($a^\dag$) destroy (create) a cavity photon, and $ \sigma_{j,j^\prime}$ represents the transition  operator from the $|j^\prime\rangle$ to $|j\rangle$ state.
In the above equation the first term in the square bracket describes the interaction of the transition X-XX with cavity photons, being $2g$ single-photon Rabi frequency, ${\cal E}_{\rm c}(t)$ is the control optical field driving resonantly the transition G-X, while ${\cal E}_{\rm in}(t)$ describes the input field feeding the microcavity.  In all subsequent calculations we will use for the 
single-photon Rabi frequency $2g = 200$ $\mu$eV \cite{Imamoglu} and a XX binding energy $\Delta = \omega_{\rm xx} - 2 \omega_{\rm x} = 4$ meV.
Losses can be taken into account within the quantum
Master equation in Born-Markov approximation for the
system density matrix $\rho$, which is expressed in the usual
Lindblad form \cite{Mandel}. The relevant Master equation for this
model is
\begin{equation}\dot{\rho}= i \left[\rho,H\right] - \frac{1}{2}\sum_{\mu} \left( L^\dag_\mu  L_\mu \rho + \rho  L^\dag_\mu  L_\mu
 -2  L_\mu \rho  L^\dag_\mu \right)\, ,
\label{rho}\end{equation}
where the Lindblad operators $ L_\mu$ describe the various decay channels: $L_c = \sqrt{\gamma_c}a$, $L_{ij} = \sqrt{\gamma_{ij}}\sigma_{ j, i}$. We employ zero temperature reservoirs.
In all subsequent calculations we will use typical decay rates for state of art QD- microcavity systems \cite{Imamoglu}. The decay rate for cavity photons is $\gamma_{\rm c} = 70$ $\mu$eV, the decay rate of the transition ${\rm XX \to X}$ is $\gamma_{\rm XX \to X} =30$ $\mu$eV, and that associated to the ${\rm X \to G}$ transition is  $\gamma_{\rm x \to g} = 20$ $\mu$eV. The influence of pure dephasing noise \cite{Auffeves} on the quantum effects under investigation can be addressed including a further term on the right-hand side of  Eq.\ \ref{rho}:
$${\cal L}_{\rm d} = - \gamma^{\rm (d)}_1 \sigma_{\rm x, x}\, \rho\, \sigma_{\rm g, g} - \gamma^{\rm (d)}_2 \sigma_{\rm xx, xx}\, \rho\, \sigma_{\rm x, x} + {\rm H.c.}$$
Starting from the above master equation, we may derive the coupled equations of motion for the
cavity-photon and exciton populations, coherences and higher order correlation functions, which we solve by representing the photon operators on a basis of Fock number states.

According to the input-output formulation of optical cavities, the external field is related to the intracavity field by the  relation \cite {walls, gardiner}
\begin{equation}
 a_{\rm out}=a_{\rm in} - \sqrt{\gamma^\prime_{\rm c}} a\, .
 \end{equation}
Here $\gamma^\prime_{\rm c}$ describes the loss through the output port. In the absence of other photon losses $\gamma^\prime_{\rm c}=\gamma_{\rm c}$.
If the input field at the output port is in the vacuum state, the output photon rate
is proportional to the intracavity photon number: $\langle a^{\dag}_{\rm out} a_{\rm out}\rangle=\gamma^\prime_{\rm c}\langle a^{\dag}a\rangle$.
An input output relationship can also be derived for a point-like QE emitting light in free space or in the absence of optical feedback.
The outward propagating part of the field at position ${\bf r}$ and time $t$ can be written as \cite {walls}:
\begin{equation}
{\bf E}_{\rm out}^{+}({\bf r},t)={\bf E}^{+}_{\rm in}-{\mathbf \Psi}(\bf r)\sigma_-(\bar{t})\, ,
\end{equation}
where ${\bf E}^{+}_{\rm in}$ is the input field, $\sigma_-$ is the QE lowering operator,  $\bar{t} = t-{\bf r}/c\approx t$, and ${\mathbf \Psi}$  is the  retarded field generated by a point dipole and it is proportional to the dipole moment associated to the emitter.
Hence the emitted light intensity is proportional to
\begin{equation}
\langle E_{\rm out}^{-}(t) E_{\rm out}^{+}(t)\rangle=|\Psi(\bf r)|^2\langle \sigma_+(\bar{t})\sigma_-(\bar{t})\rangle\, .
\end{equation}
In the present case, considering the light field emitted by the exciton level X, we have $\sigma_- = \sigma_{\rm gx}$.

\section{Numerical Results}
Let us consider the time resolved response  after the arrival of fast optical pulses.
At initial time,  a weak Gaussian control pulse (resonant with the transition G-X; pulse area = $0.012 \pi$) performs a very small rotation of the Bloch sphere  $\{| {\rm G} \rangle,\, | {\rm X} \rangle \}$ giving rise to a superposition $| \psi(t) \rangle = |0 \rangle ( a_{\rm g}| {\rm G} \rangle + a_{\rm x} | {\rm X} \rangle)$ with $| a_{\rm x} |^2 << | a_{\rm g} |^2$. Immediately after (3 ps delay) a weak probe beam (pulse area = $0.02 \pi$) resonant with the bare cavity-mode feeds the microcavity. The time-duration (FWHM) of both pulses is 5 ps.
In Figure \ 2, we present our numerical results for experimentally accessible observables of the system.
In Figure \ 2a we plot the time dependence of the cavity average photon number $\langle a^\dag a \rangle$ which is proportional to the detectable output transmitted photon flux
$\langle a^{\dag}_{\rm out} a_{\rm out}\rangle=\gamma^\prime_{\rm c}\langle a^{\dag}a\rangle $ if the input field at the output port is in the vacuum state.
It displays irreversible exponential decay mainly due to photon escape through the cavity mirrors,  which is a signature of weak (or even absence of) light-matter interaction.
Observation of the vacuum Rabi splitting or, in the time domain, vacuum Rabi oscillations,  requires that the effective coupling rate
(known to scale with the electronic population difference among the involved levels) exceeds the mean of the decay rates of the cavity and of the electronic transition. In a semiclassical picture, the effective coupling rate is given by $\tilde g = g \sqrt{|\langle \sigma_{\rm x, x}\rangle- \langle \sigma_{\rm xx, xx} \rangle|}$.  In the present case, owing to the smallness of $\langle \sigma_{\rm x, x} \rangle$ as well as of $\langle \sigma_{\rm xx, xx} \rangle$, the system displays a weak coupling regime behavior. The superimposed small and fast oscillations originate from the coupling of the cavity with the detuned X-G transition.
Figure\ 2 also shows the level $| {\rm X} \rangle$  population dynamics which also exhibits irreversible Markovian behavior, due to spontaneous emission.
Spontaneously emitted photons at frequency $\omega_{\rm x}$ radiated out the side of the cavity, are proportional to
$\langle \sigma_{\rm x,x } \rangle$. Hence it is possible to gather direct information on both the cavity-photon population  (by detecting photons escaping from one cavity mirror at $\omega_{\rm c} = \omega_{\rm xx} - \omega_{\rm x}$ and on $\langle \sigma_{\rm x, x}  \rangle$.
In Figure \ 2a we also plot the time behavior of the joint probability of detecting one cavity photon and one photon from the spontaneous transition $| {\rm X} \rangle \to |{\rm G} \rangle$ which is proportional to $I_{\rm xc} = \langle a^\dag a\, \sigma_{\rm x,x}\rangle$. This coincidence rate, a subset of the Markovian signals
$\langle \sigma_{\rm x, x}\rangle$ and $\langle a^\dag a \rangle$,  displays almost perfect vacuum Rabi oscillations.

It is worth noticing that the detected joint signal will actually be proportional to the joint output field intensities $\langle a_{\rm out}^\dag a_{\rm out}E^-_{\rm out}
E^+_{\rm out}\rangle$, which in turn is proportional to  $I_{\rm xc} = \langle a^\dag a\, \sigma_{\rm x,x}\rangle$. Hence the actually measured joint dynamics is correctly described by $I_{\rm xc}$.
We also observe that the cavity is efficient and emits the photons directionally (in the cavity axis), while the emitter is long-lived, emits weakly, and isotropically. As a consequence the experimental observation of such a joint detection rate will require  much longer acquisition times than those required for the acquisition of
$\langle a_{\rm out}^\dag a_{\rm out} \rangle$.
Figure 2a also displays $I_{\rm xc}$ calculated including additional pure dephasing. We adopted $\gamma^{\rm (d)}_1 =15$  $\mu$eV and $\gamma^{\rm (d)}_2 = 20$ $\mu$eV. The observed effect reveals robust against a significant amount of pure dephasing which just reduces the amplitude of vacuum Rabi oscillations.
This is an interesting example of correlated quantum dynamics induced by the quantum superpositions of on/off strong coupling between a microcavity and a single QD. The optical observables here addressed, display nonclassical behavior, as can be inferred from the violation
of the Cauchy-Schwarz inequality for classical correlations \cite{Mandel}:
$$[g_{\rm c x}^{(2)}(t,t)]^2 \leq g_{\rm c}^{(2)}(t,t) g_{\rm x}^{(2)}(t,t),$$ where  $g_{\rm c x}^{(2)}(t,t) = I_{\rm xc}/( \langle a^\dag a \rangle \langle \sigma_{\rm x, x} \rangle)$ is the zero-delay two-mode normalized second-order intensity correlation function and $g_{\rm c (x)}^{(2)}(t,t)$ are the corresponding single-mode normalized second-order correlation functions for cavity photons and photons emitted from the spontaneous decay of level X, respectively.

In order to explore more deeply such quantum correlated dynamics, we calculated the joint probability of detecting a cavity photon at $t_1$ and
one photon from the decay of level X at $t_2$  (see Fig.\,  3a):
$$I_{\rm c x}(t_1, t_2) = \langle {\cal T} a^\dag (t_1) \sigma_{\rm x, x} (t_2) a (t_1) \rangle\, .$$
The symbol ${\cal T}$ denotes time ordering: {e.g.} if
$t_1 > t_2$, $$I_{\rm c x}(t_1, t_2) = \langle \sigma_{\rm x,g}(t_2) a^\dag (t_1)  a (t_1) \sigma_{\rm g,x} (t_2)\rangle\, .$$
The figure shows a clear coexistence of exponential decay an Rabi oscillations.
 Such coexistence is a signature
of the ``suspended'' dynamics  between two completely different time evolutions induced by the quantum superposition $| \psi(t)\rangle = a | {\rm On}(t) \rangle + b | {\rm Off}(t) \rangle$.
Such behavior can be better understood looking at line cuts obtained fixing $t_{1} = \bar t_{1}$ or $t_{2} = \bar t_{2}$.
For example the functions ${\cal C}_{\bar t_2}(t) =  I_{\rm c x}(t, \bar t_2)$
displays an oscillatory behavior for $t< \bar t_2$ and an exponential decay for $t > \bar t_2$. An analogous behavior is exhibited by
${\cal X}_{\bar t_1}(t) =  I_{\rm c x}(\bar t_1, t)$. Hence the presence of a subsequent detection event affects the previous photonic dynamics.
On the other hand a detection event {\em e.g.} at time $\bar t_2$ determines a collapse of the wavefunction affecting the dynamics for the following times $t > \bar t_2$.
	Figure \, 3b shows the normalized joint probability of detecting a cavity photon at $t_1$ and
	one photon from the decay of level X at $t_2$:
	$$P^N_{12}(t_1, t_2)=\frac{I_{\rm c x}(t_1, t_2)} { \langle  a^\dag (t_1)a (t_1)\rangle \langle \sigma_{\rm x, x} (t_2) \rangle}\, .$$
	We observe that at increasing times   $P^N_{12}$ increases. This is due to the fact that increasing $t_1$ and $t_2$ the X-state depopulates and all terms composing $P^N_{12}(t_1, t_2)$ goes to zero. $I_{\rm c x}(t_1, t_2)$ is an infinitesimal of same order as $\langle \sigma_{\rm x, x} (t_2) \rangle$, hence the further decaying term
		$\langle  a^\dag (t_1)a^\dag (t_1)\rangle $ in the denominator  causes the increasing of   $P^N_{12}$ shown in Fig.\, 3b at greater times. 
	Figure 3c displays the cavity photon number $\langle a^\dag a \rangle$ as well as the coincidence rate $I_{\rm xc} = \langle a^\dag a\, \sigma_{\rm x, x} \rangle$ under  continuous-wave excitation as a function of the frequency of the input field feeding the microcavity ${\cal E}_{\rm in}=0.003$.
	  In this calculation we will use for the 
	  single-photon Rabi frequency $2g = 100$ $\mu$eV and a XX binding energy $\Delta = \omega_{\rm xx} - 2 \omega_{\rm x} = 4$ meV. The system is also excited with a weak monochromatic field control pulse ${\cal E}_{\rm c}=0.001$ driving resonantly the transition G-X. While $\langle a^\dag a \rangle$ displays a single peak at the cavity-mode energy, characteristic of a system in the weak-coupling regime, $I_{\rm xc}$ exhibits the vacuum Rabi splitting.
Results in Fig.\, 3c demonstrate that the  conditional quantum dynamics induced by the quantum superposition $| \psi(t)\rangle = a | {\rm On}(t) \rangle + b | {\rm Off}(t) \rangle$ is  also present in the  nonequilibrium steady state of this coherently driven dissipative quantum systems.
\begin{figure}
	\centering
	\includegraphics[width=0.9\textwidth]{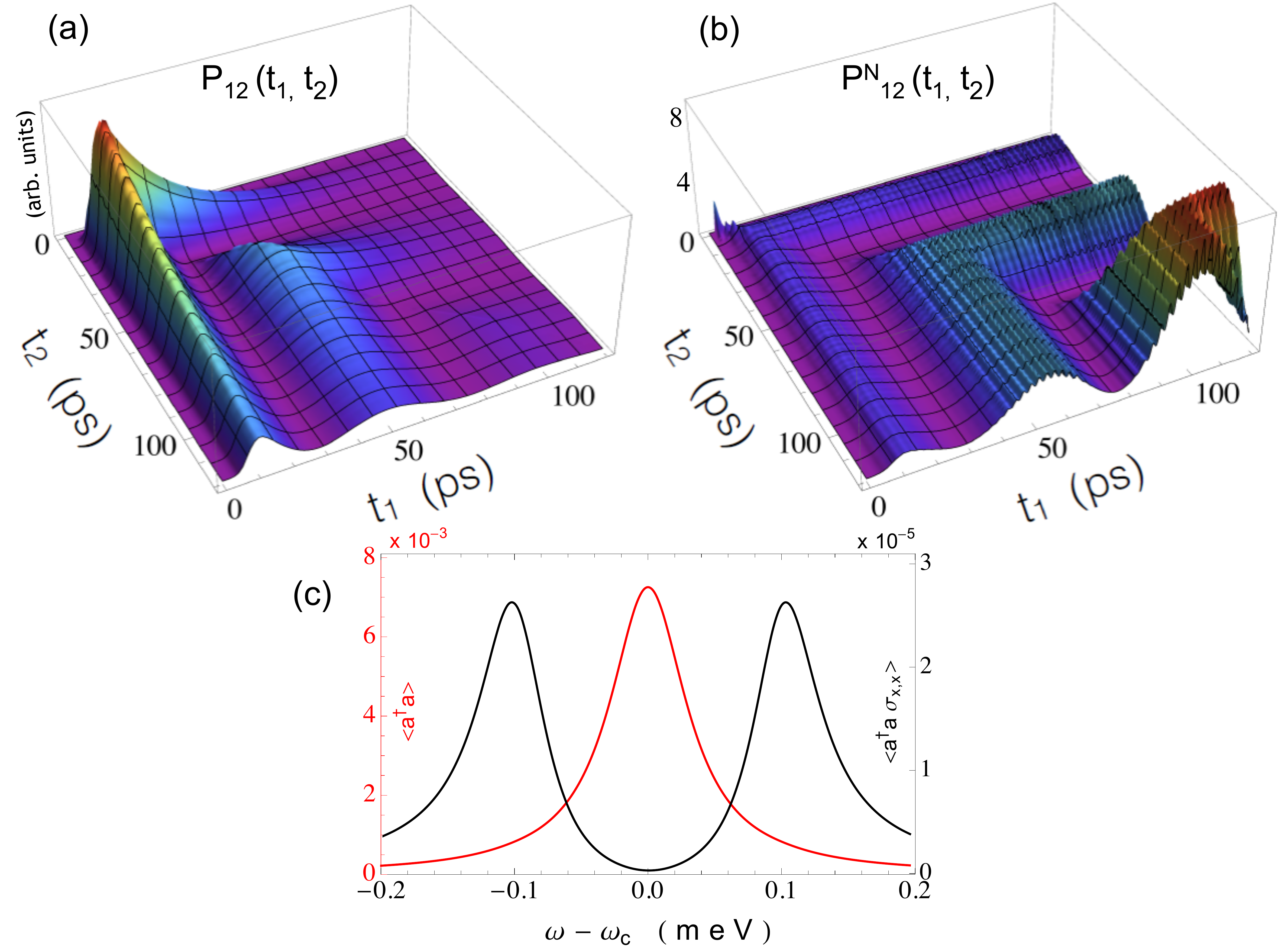}
	\caption{(Color online). (a)  Joint probability of detecting a cavity photon at $t_1$ and
		one photon from the decay of level X at $t_2$: $P_{1,2}(t_1, t_2)=I_{\rm c x}(t_1, t_2)$.
		(b) Normalized joint probability of detecting a cavity photon at $t_1$ and
		one photon from the decay of level X at $t_2$: $P^N_{1,2}(t_1, t_2)=I_{\rm c x}(t_1, t_2)/\langle a^\dag a\rangle \langle \sigma_{\rm gx} ^\dag \sigma_{\rm gx}\rangle $
		after two pulse excitation as in Fig.\ 2.
		(c) Cavity photon number $\langle a^\dag a \rangle$ and coincidence rate $I_{\rm xc} = \langle a^\dag a\, \sigma_{\rm x,x} \rangle$ under  continuous-wave excitation as a function of the frequency of the input field feeding the microcavity ${\cal E}_{\rm in}$. The system is also excited with a weak monochromatic field ${\cal E}_{\rm c}$ driving resonantly the transition G-X. } \vspace{-0.3 cm}
\end{figure}

\section{Conclusions}
We have theoretically investigated the time- and frequency-resolved spectroscopic signatures  of quantum superpositions
of on/off strong coupling between a microcavity and a single QD
The resulting system dynamics, ``suspended'' between  two different time evolutions, has been discussed using readily accessible
optical observables.
The experimental preparation and observation of the proposed dynamics in a single QD-semiconductor microcavity system is within experimental
reach with existing state-of-the-art technology \cite{Bochmann,Moreau}. Possible experimental issues can originate from the excitonic emission
which is not directional and can be weak with respect to the cavity emission. However, since only coincidence events are recorded in
such (coincidence) experiments, attenuation of the stronger signal or low efficient collection of the other signal does not alter
the normalized cross-correlation functions. It determines only an increase of the events not displaying coincidences which will be
discarded (post selection), resulting just into an increase of the acquisition times.
An interesting example of cross correlation measurements in a cavity embedded single QD has been reported in Ref. \cite{Imamoglu}.
The effects here described  can also be  studied  in  novel ultracompact   hybrid structures composed by a single QD strongly
coupled via localized surface plasmons to one or a couple of metallic nanoparticles \cite{Ridolfo2010, Savasta2010}.
Laser cooling of the center-of-mass
motion of such hybrid system has been recently proposed  \cite{Marago}. Realization of On-Off interaction superpositions in these mesoscopic
systems can result into  quantum superpositions of  motional  states.
The scheme here proposed, based on tuning the mode energy of the microcavity to the biexciton-exciton (XX-X) transition of the QD,
could be exploited for the solid state implementation of scalable quantum computation
with single-photon polarizations as qubits  \cite{PRLKimble}.
It would
also be interesting to extend such quantum control in
optically active many-body quantum systems \cite{OpticalBook,Portolan2009,DiStefano2001,Portolan2008,Portolan20082}
and in  quantum optical systems in the ultrastrong coupling regime \cite{Stassi2013,Garziano2015,Garziano2016, DiStefano2016}.

\bibliographystyle{elsarticle-num}
\bibliography{catnew}

\newpage
%
%


\newpage

\end{document}